 \documentclass[final,3p,times]{elsarticle}

\usepackage{amssymb}
\usepackage{amsmath}

\usepackage{graphicx}
\usepackage{bm}

\journal{Chinese Journal of Physics}

\begin{document}

\begin{frontmatter}



\title{Sparse Edge Encoder (SEE): I. Visual recognition in neuronal networks}

\author[1]{Chia-Ying Lin\fnref{fn1}}
\author[1]{Mei Ian Sam\fnref{fn1}}
\author[1]{Yi-Ching Tsai}
\author[1,2]{Hsiu-Hau Lin\corref{cor1}}

\cortext[cor1]{Correspondence and requests for materials should be addressed to Hsiu-Hau Lin (hsiuhau.lin@phys.nthu.edu.tw).}
\fntext[fn1]{These authors CYL and MIS contribute equally.}

\affiliation[1]{organization={Department of Physics},
addressline={National Tsing Hua University},
city={Hsinchu},
postcode={300044},
country={Taiwan}}

\affiliation[2]{organization={College of Semiconductor Research},
addressline={National Tsing Hua University},
city={Hsinchu},
postcode={300044},
country={Taiwan}}

\begin{abstract}
In the past few decades, there have been intense debates whether the brain operates at a critical state. To verify the ``criticality hypothesis" in the neuronal networks is challenging and the accumulating experimental and theoretical results remain controversial at this point. Here we simulate how visual information of a nature image is processed by the finite Kinouchi-Copelli neuronal network, extracting the trends of the mutual information (how sensible the neuronal network is), the dynamical range (how sensitive the network responds to external stimuli) and the statistical fluctuations (how criticality is defined in conventional statistical physics). It is rather remarkable that the optimized state for visual recognition, although close to, does not coincide with the critical state where the statistical fluctuations reach the maximum. Different images and/or network sizes of course lead to differences in details but the trend of the information optimization remains the same. Our findings pave the first step to investigate how the information processing is optimized in different neuronal networks and suggest that the criticality hypothesis may not be necessary to explain why a neuronal network can process information smartly.
\end{abstract}




\end{frontmatter}



\section{Introduction}
\label{sec1}
Vision is essential for all animals to interact with their environment. In humans, visual perception is mediated by complex mechanisms that span from the eyes to the brain\cite{Galizia13}. Visual recognition, therefore, arises from the interplay between external stimuli and the neural systems that process them\cite{Gonzalez18,Kuehni08}. A recent patent, Sparse Edge Encoder (SEE), inspired by principles from physics and neuroscience, reveals hidden statistical patterns in natural images to optimize image data encoding and decoding\cite{Lin24,Lin25}. To elucidate the microscopic mechanisms of the SEE algorithm, we will present a series of three papers, beginning with an exploration of visual recognition in neuronal networks here. In the second paper, we will examine color perception through the lens of symmetry breaking and identify the optimal grayscale axis (the symmetric axis) for color encoding. Finally, in the third paper, we will integrate these insights with the SEE algorithm to develop an optimized framework for encoding and decoding color images.

Back in the 1990's, Dunkelmann and Radons\cite{Dunkelmann94} showed that the neuronal networks exhibit self-organized criticality. Later, experimental works by Beggs and Plenz\cite{Beggs03} further promoted the ``critical hypothesis" for optimized information processing in the brain. However, even after decades of intense studies\cite{Wilting19,OByrne22,Presigny22,Janusz18}, the hypothesis remains controversial. Being critical enjoys several benefits. First of all, the critical neurons respond sensitively to various external stimuli. The physical origin of the enhanced responses arises from various divergent susceptibilities at the critical point. Meanwhile, there are considerable experimental observations of the power-law behaviors. Avalanches of neuron firing in various regimes of the brains (both in human and other animals) are found to follows power-law distributions via different probes such as MEG, EEG and BOLD signals\cite{Mazzoni07,Pasquale08,Tetzlaff10,Friedman12,Priesemann09,Petermann09,Yu17,Hahn10,Hahn17,Hansen01,Shriki13,Palva13,Tagliazucchi12}. These evidences provide solid supports for the critical hypothesis. 

Because the brain is complex\cite{Presigny22,Rossa20,Engel21,Danziger22}, it is not surprising that somehow the system self-organizes itself into some critical states, generating all sorts of the observed power-law exponents. However, it does not necessarily indicate that a critical brain can better process the information. One shall return to the core question whether the optimized information processing\cite{Beggs08,Tanaka09,Shew11,Shriki13,Shriki16,Dettner16,Takagi20} is related to the critical dynamics. While the information processing in the biological brains is not yet fully understood, optimizing the mutual information in machine learning\cite{Belghazi18,Hjelm19,Kong20} has led to tremendous success in the past few years. 

However, a deep look into the hypothesis brings up confusions as well. The phenomena of the so-called critical slow down\cite{Scheffer12} is notorious and the time needed to complete a sensible information processing can be extremely long. In fact, the time scale diverges right at the critical point. Besides, the enhanced responses in the unwanted aspects can be negative, leading to reduced specificity and reliability\cite{Gollo17,Wilting18}. Therefore, criticality is not always helpful if one focuses on the information processing.

In this Article, we revisit this important question, focusing on the visual recognition at the very front end of the visual neural system, such as ganglion cells in the retina. We intentionally choose a relatively simple neuronal network to avoid the unnecessary complexity. Meanwhile, we also use the pixel data generated from natural images to mimic the realistic visual information. Choosing the finite Kinouchi-Copelli neuronal network (KCNN)\cite{Kinouchi06,Lin22} to simulate how visual information of a nature image is processed, we successfully extract the trends of the mutual information, the dynamical range and the statistical fluctuations altogether. As will be elaborated in later paragraphs, the optimized state of the KCNN for visual recognition does not coincide with the critical state where the statistical fluctuations reach the maximum. Further analysis reveals that this deviation is robust, casting shadow on the critical hypothesis. Because the KCNN is relatively simple, the critical state can be easily spotted even in the finite system. Our numerical simulations suggest that the critical hypothesis may not be necessary and the optimized information processing is due to other competing mechanism. In the following, we would like to establish the maximum mutual information principle first and move on the our numerical simulation in the KCNN.

\section{Maximum mutual information principle}
\label{sec2}

Let us briefly review how information is processed in a neuronal network\cite{Haykin09,Goodfellow16}. Consider the general structure shown in Figure 1. The multichannel input $\bm{X}= X_1, X_2, \cdots X_n$ is processed by the neuronal network, generating the multi-channel output $\bm{Y}=Y_1, Y_2, \cdots, Y_m$. If we are interested in extracting the input information from the output signals, one shall minimize the conditional information entropy $H(\bm{X}|\bm{Y})$ by data training or hyperparameter optimization in the neuronal network. That is to say, once the output signals are known through measurements, the uncertainty to extract the input information in minimized. 

\begin{figure}
\begin{center}
\includegraphics[width=\columnwidth]{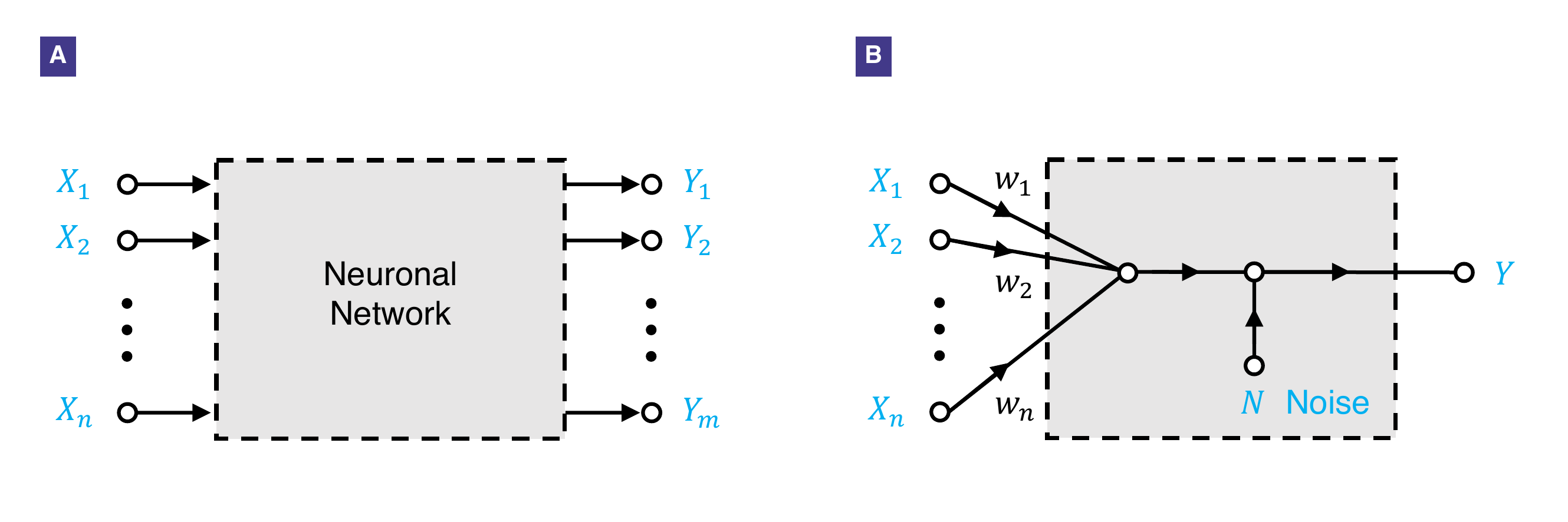}
\caption{\textbf{Information processing in the neuronal network.} (A) The neuronal network processes the multichannel input $\bm{X} = (X_1,X_2,\cdots,X_n)$ and generates the multichannel output $\bm{Y} = (Y_1,Y_2,\cdots,Y_m)$. Note that the numbers of the input and output channels are not necessarily the same. (B) The single linear neuron connects with the inputs $X_i$ through the synaptic weights $w_i$ (acting as dendrite) and generates a single output $Y$ (as firing rate) with processing noise $N$. The the noise $N$ and the multichannel inputs $X_i$ are not correlated, and both are taken to be Gaussian for simplicity.}
\end{center}
\end{figure}

Making use of the identity for mutual information $I(\bm{X},\bm{Y})$, it can be expressed as the difference between the information entropy of the input $H(\bm{X})$ and the conditional information entropy $H(\bm{X}|\bm{Y})$,
\begin{eqnarray}
I(\bm{X},\bm{Y}) = H(\bm{X}) - H(\bm{X}|\bm{Y}).
\end{eqnarray}
Because the information entropy $H(\bm{X})$ is fixed (independent of the training/optimization in the neuronal network), minimizing the conditional information entropy $H(\bm{X}|\bm{Y})$ is equivalent to maximizing the mutual information $I(\bm{X}|\bm{Y})$ between the input and output signals.

To demonstrate the maximum mutual information principle, it is helpful to work out the example for the simplest neuronal network -- a single linear neuron as shown in Figure 1(B). For a neuron with the linear gain function, the input and output are related by the linear relation,
\begin{eqnarray}
Y = \sum_{i} w_i X_i + N,
\end{eqnarray}
where $w_i$ represent the synaptic weights, satisfying the normalization condition $\sum_i w_i^2 =1$. The noise $N$ describes the processing noise modeled as a Gaussian random variable with zero mean and a variance $\sigma^2_{N}$. For simplicity, we would also assume the inputs $X_1, X_2, \cdots, X_n$ are Gaussian distributed but not necessarily independent to each other.

The mutual information can be easily computed through the following relation,
\begin{eqnarray}
I(\bm{X}, Y) = H(Y) - H(Y|\bm{X}) = \frac12 \left[1+ \ln(2\pi\sigma^2_Y) \right] - \frac12 \left[1+ \ln(2\pi\sigma^2_N) \right] = \frac12 \ln \left( \frac{\sigma^2_Y}{\sigma^2_N}\right)
\end{eqnarray}
The above result agrees with our intuition arisen from the signal-to-noise ratio $\sigma^2_Y/\sigma^2_N$. For the linear neuron consider here, the output variance can be further simplified,
\begin{eqnarray}
\sigma^2_Y = \sum_{ij} w_i \langle X_i X_j \rangle w_j + \sigma^2_N.
\end{eqnarray}
In the above calculations, we have assumed that the inputs $X_i$ do not correlate with the noise $N$. It becomes clear now that maximizing the mutual information is equivalent to find the maximum eigenvalue of the covariance matrix $\langle X_i X_j \rangle$ of the inputs and the optimized synaptic weights $w_i = \arg \max I(\bm{w})$.

\section{Structured visual information}
\label{sec3}

The example of the single linear neuron is inspiring but misses several important ingredients. First of all, the network structure is important and the consequences originated from a single-neuron model can be misleading. Secondly, the linear gain function does not address the essential nonlinearity in the realistic neuronal networks. Finally, if one would like to test the ``criticality hypothesis" of the neuronal network, the parameter regime for optimization shall cover at least two different phases.

\begin{figure}
\begin{center}
\includegraphics[width=12cm]{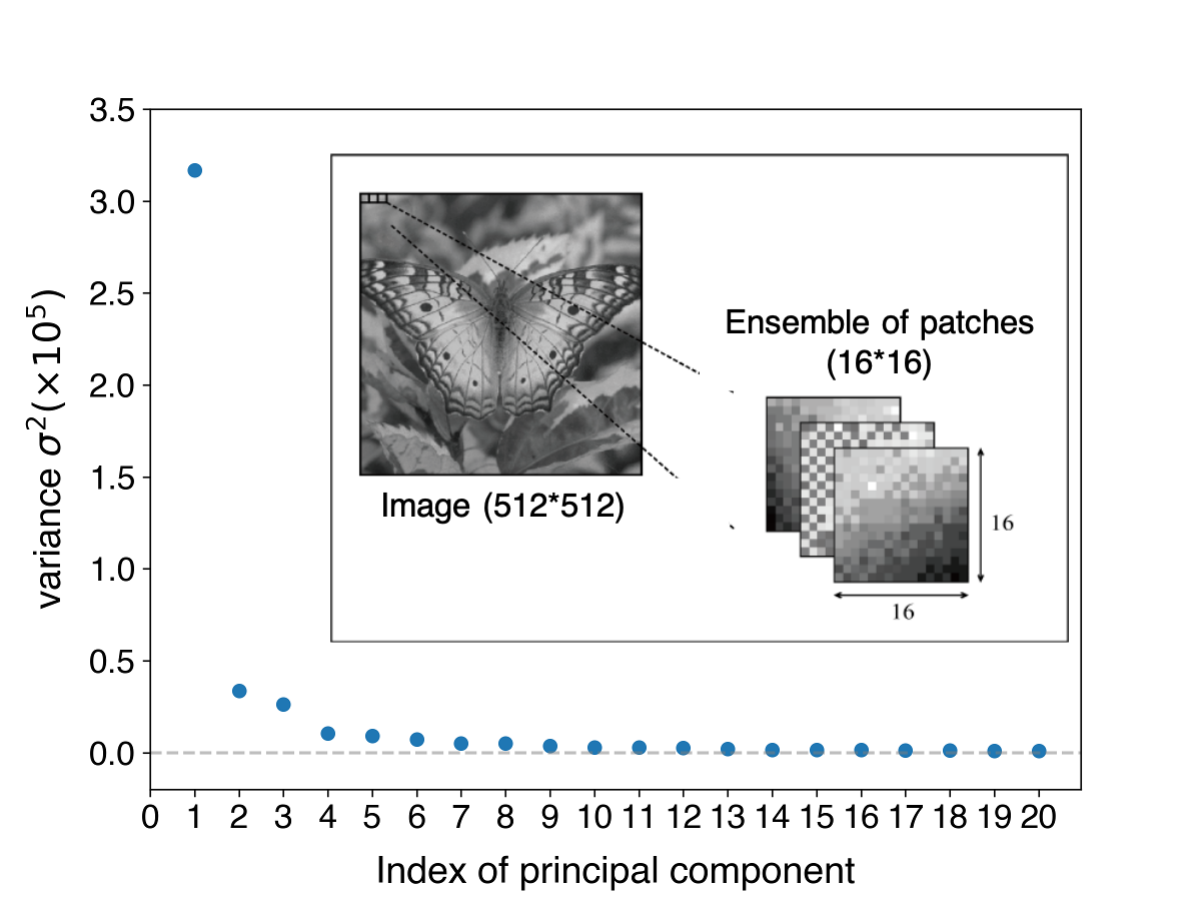}
\caption{\textbf{Principal component analysis for a natural image.} Taking a $512 \times 512$ natural images and slicing into an ensemble of $L \times L$ patches with $L=16$, one can obtain the hidden statistical patterns at this small length scale by the principal component analysis. if all $L^2=256$ pixels are randomly distributed, all eigenvalues (variances) of the principal components are degenerate. In the opposite limit where all pixels are perfectly correlated, the spectrum of the eigenvalues is the Dirac delta function. It is clear that the eigenvalue spectrum of a natural image (butterfly here) takes the subtle balance between these two extreme limits.}
\end{center}
\end{figure}

To put our feet on solid ground, we investigate how visual information in processed in a finite neuronal network. Even at the very front end of our visual systems, the inputs are not encoded pixel by pixel, as evident in the discovery of the receptive field. Thus, we adopt a finite $L \times L$ neuronal network as the base for visual information detection. As elaborated in the previous work, the U(1) dynamics is universally found in most single-neuron models and the Kinouchi-Copelli neuron can be viewed as a discrete version of the U(1) neuron. Technical details of the Kinouchi-Copelli neuron can be found in the Supplementary Information.

Before diving into the details of numerical simulations, it is important to highlight the intrinsic structures within natural images. As shown in Figure 2, the image ``butterfly" of the size $512 \times 512$ is cut into an ensemble of $16 \times 16$ patches. Treating each patch as a $256$ dimensional input vector, one can extract the prioritized statistical patterns by the principal component analysis (PCA)\cite{Haykin09,Goodfellow16}. As expected, at the small scale ($16 \times 16$), the first principal component corresponds to the near-uniform eigenvector with the largest eigenvalue. The next few principal components correspond to variations at the small length scale, characterized with increasing numbers of the nodal lines. Typically, one can take roughly 15\% of the principal components of a natural image (disregarding all other principal components) and the synthesized image appears identical to human eyes.

While this is known in image analysis for decades, it is worth emphasizing that the PCA shows that the visual information of a natural image at the small length scale is structured. If all $256$ inputs are independent random variables (generated by computer simulations), the eigenvalues obtained by the PCA would all be degenerate. On the other hand, if all inputs are perfectly correlated, the eigenvalue spectrum turns into the Dirac delta function: except that of the first principal components, all other eigenvalues are zero. Apparently, a natural image contains structured visual information and does not follow either extreme limits.

\section{Mutual information, dynamical range and statistical fluctuations}
\label{sec4}

Here we employ the Kinouchi-Copelli neuronal network (KCNN) of the size $16 \times 16$ to simulate how visual information is processed at the front end of the visual neural system (such as ganglion cells in the retina). As summarized in Figure 3, we use Monte Carlo simulations to compute the mutual information (how sensible the KCNN is), the dynamical range (how sensitive the KCNN is) and the statistical fluctuations (the conventional indicator for criticality). Sampling from the natural image ``butterfly", one builds up an ensemble of $16 \times 16$ patches to serve as the incoming stimuli $S_i$, where $i=1,2,\cdots,L^2$ with $L=16$. To capture the spatial variations at the small length scale, the inputs are defined as
\begin{eqnarray}
X_i = S_i - \langle S \rangle = \frac{1}{L^2}\sum_{j}^{L^2} S_j.
\end{eqnarray}
Note that the $i$-th neuron in the KCNN can be activated by the external stimuli (the input $X_i$) and/or the synaptic interaction with its neighbor neuron $j$ (the synaptic weight $W_{ij}$). The resultant activity of the $i$-th neuron is then captured by the output $Y_i$. By varying the synaptic weights $W_{ij}$, one seeks for the optimized parameters where the mutual information between the inputs and outputs is maximized.

For simplicity, we consider the KCNN with uniform network structure, i.e. $W_{ij}=W/4$, where the factor of 4 arisen from the number of neighboring neurons. As explained in the Supplementary Information, the relation between the branching ratio $\sigma$ and the synaptic weight $W$ is derived. For the uniform KCNN, the optimization problem is greatly simplified. Even though the uniform network is a simplification, it still contains the essential ingredient for driving the phase transition from the quiescent phase (Q phase) to the spontaneous asynchronous firing phase (SAF phase).

\begin{figure}
\begin{center}
\includegraphics[width=\columnwidth]{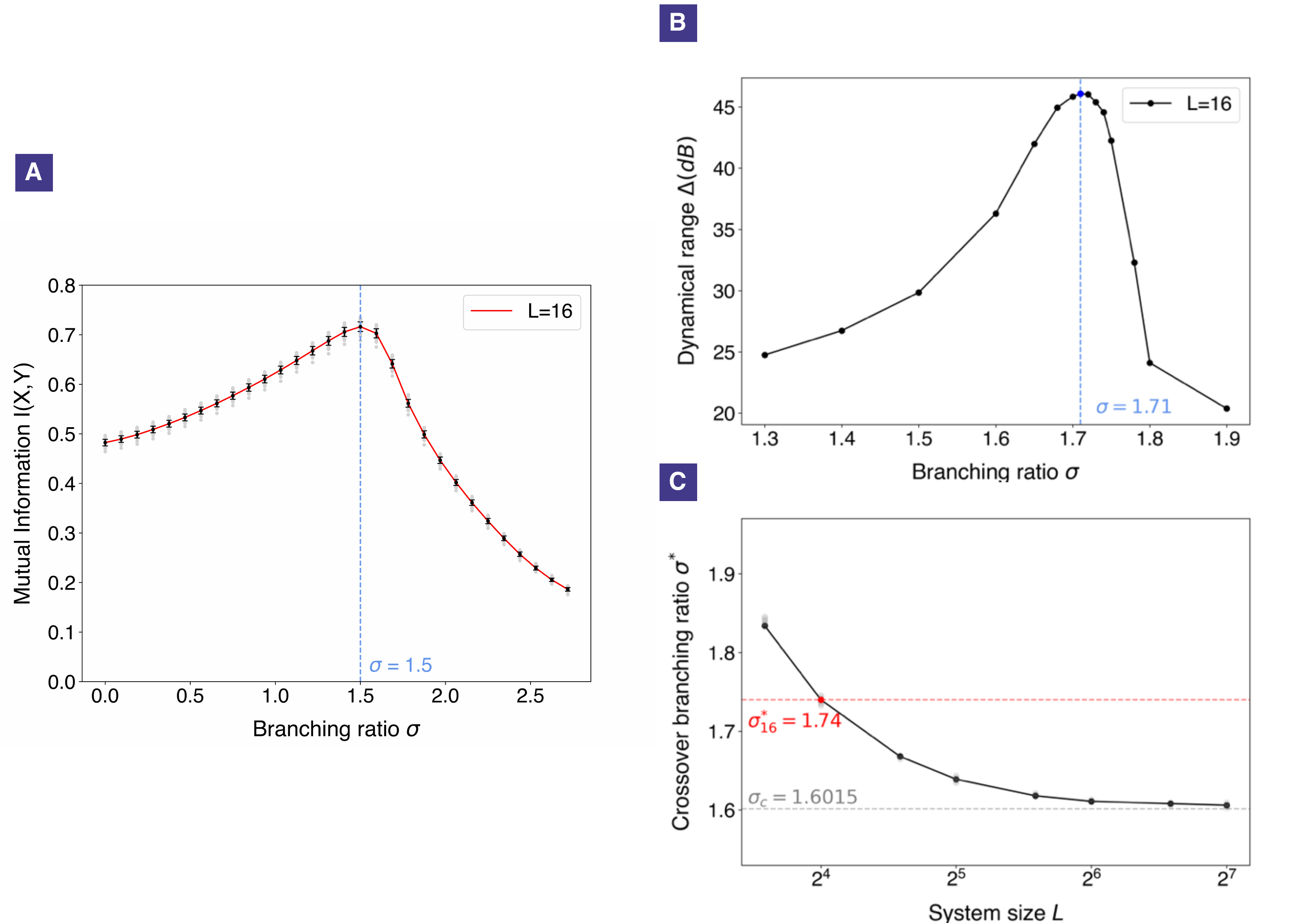}
\caption{\textbf{Sense and sensibility of a neuronal network.} (A) The mutual information for visual recognition in a $16 \times 16$ KCNN increases gradually as the branching ratio increases and reaches the peak around $\sigma_{\rm MI} = 1.5$. Further increase of the branching ratio beyond $\sigma_{\rm MI}$ does not help, leading to a quick drop in the mutual information. That is to say, the optimized information processing requires a subtle balance of the synaptic interactions. (B) The dynamical range measures how sensitive the neuronal network reacts to the external stimuli. Our numerical simulations show that the dynamical range reaches its maximum around $\sigma_{\rm DR} = 1.71$, quite close to the crossover $\sigma^*= 1.74$ where the statistical fluctuations are the strongest. (C) The crossover branching ratio $\sigma^* = \sigma^*(L)$ for the KCNN is defined as the onset where the statistical fluctuations reach the maximum. The finite-size scaling (shown in the Supplementary Information) leads to the critical branching ratio $\sigma_c = 1.6015$ in the thermodynamic limit.}
\end{center}
\end{figure}

Strictly speaking, a critical point is only well defined in the thermodynamic limit. For comparison, the fluctuations of the activity in the KCNN are computed at various branching ratios and the crossover $\sigma^*$ is introduced as the onset of maximum statistical fluctuations. We have carefully checked the finite-size scaling and extracted the critical branching ratio $\sigma_c = \lim_{L\to\infty} \sigma^*(L) = 1.6015$. For the system size of $L=16$, the crossover branching ratio $\sigma^* = 1.74$.

Is the maximum mutual information achieved at the same branching ratio $\sigma^*$? The answer is No. As shown in Figure 3(A), as the branching ratio increases gradually, the mutual information $I(X,Y) = I(\sigma)$ increases slowly at the beginning, reaching the maximum at $\sigma_{\rm MI} = \arg\max I (\sigma) = 1.5$. After passing the peak, the mutual information drops significantly to small values, implying low efficacy to process visual information. When compared with the crossover branching ratio $\sigma^*$, it is clear that $\sigma_{\rm MI}$ is within the Q phase where no spontaneous firing activity occurs.

One can also look at the dynamical range, a measure of how sensitive the neuronal network responds to the external stimuli. As shown in Figure 3(B), our numerical simulations show that the dynamical range reaches its maximum around $\sigma_{\rm DR} = 1.71$, quite close to the crossover branching ratio $\sigma^*=1.74$ where the statistical fluctuations reach the maximum. Note that the dynamical range can be viewed as a non-linear version of the susceptibility at the wider scope. It is thus anticipated that the peaks of the dynamical range (how sensitive the networks is) and the statistical fluctuations (how criticality is defined) are close to each other.

\begin{figure}
\begin{center}
\includegraphics[width=\columnwidth]{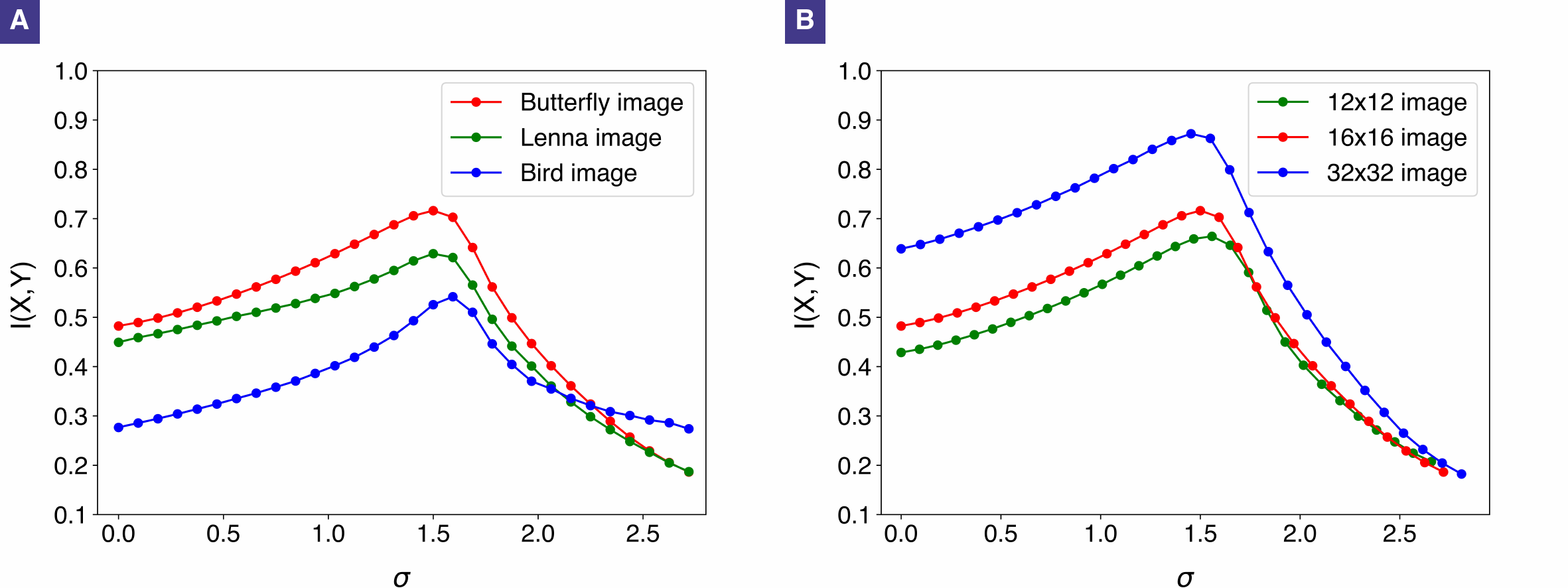}
\caption{\textbf{The same trend of the mutual information with different details.} (A) The analysis is repeated for different natural images (butterfly, Lena and bird). The trend of the mutual information remains the same with different details. The optimized branching ratios $\sigma_{\rm MI}$ for different images are slightly different, hinting its non-universal property. (B) The trend of the mutual information remains robust when the size of the neuronal network changes.}
\end{center}
\end{figure}

To further strengthen that the optimized branching ratio $\sigma_{\rm MI}$ differs from the crossover branching ratio $\sigma^*$, we repeat the same analysis for different natural images (butterfly, Lena and bird) as shown in Figure 4(A). While the details are different, the trend of the mutual information remains robust. The optimized branching ratios for different images differ slightly but a sizable deviation from the crossover branching ratio persists. We also change the size of the neuronal network and study its consequences. As shown in Figure 4(B), the mutual information for the KCNN of different sizes again shows the same trend. 

Our findings here echoes with recent studies, revealing the dynamics of the neuronal network is not critical but reverberating. Instead of maximizing the susceptibilities (statistical fluctuations), the optimized branching ratio $\sigma_{\rm MI}$ may arise from the balanced competitions between various factors like quality of representation (small branching ratio) vs integration time (large branching ratio). All these results support the conclusion that the optimized branching ratio for information processing is not universal and thus unlikely directly related to the enhanced statistical fluctuations.

\section{Discussions and open questions}
\label{sec5}

The peak structure of the mutual information can be understood in the following way. In the regime with small branching ratios, the firing activities of each neurons in the network are more or less independent so that the spatial resolution is kept. On the other hand, in the regime with large branching ratios, the neuronal activities are greatly enhanced due to synaptic interactions and the temporal distributions of the firing events become more reliable. When analyzing the visual information (with intrinsic spatial correlations) studied here, the neuronal network needs to balance the pros and cons, leading to the peak structure obtained by numerical simulations.

The key is to further differentiate the onsets of maximum statistical fluctuations and the optimized information processing. As emphasized previously, the critical point is well defined only in the thermodynamic limit, i.e. the infinite size. One thus may challenge that all comparisons in the finite-size neuronal networks are meaningless. However, all functional information processors in biological organs are finite in sizes. If the critical hypothesis only stands in the thermodynamic limit, the debate becomes moot.

More studies are required to verify whether our primitive results here are on the right track. For instance, one can choose the random inputs without any hidden statistical structures and compare with current results. Because the eigenvalue spectrum of the PCA would be drastically different, it is anticipated that the optimized branching ratio $\sigma_{\rm MI}$ would be different from that obtained from the natural images.

One can also include the spatial variations of the synaptic interactions. The uniform KCNN studied here is just a simple example. It would be exciting to explore how the mutual information is optimized when all synaptic weights $W_{ij}$ can be adjusted dynamically. Instead of exploring the entire parameter space of the synaptic weights, it makes sense to employ the machine-learning approach to train these weights $W_{ij}$ by the ensemble of the image patches.

\section*{Acknowledgement}
HHL proposed the research topic and supervise the project. CYL and MIS both performed the numerical simulations on the KCNN and finished the calculations on mutual information, dynamical range and statistical fluctuations. YCT performed the PCA for the natural images. CYL, MIS and HHL discussed the results and prepared the manuscript together. We acknowledge supports from National Science and Technology Council in Taiwan through Grant No. NSTC 113-2112-M-007-017. We are grateful to Po-Chung Chen for providing the abundant computational resources to run the numerical simulations.

\appendix
\section{Methods}
\label{app1}

The Kinouchi-Copelli model\cite{Kinouchi06} proposed in 2006 gives a concrete realization that a network of excitable elements maximizes its sensitivity (and thus dynamic range) at the critical point of a non-equilibrium phase transition. In the Article, we adopt a square-lattice network composed of the 4-state Kinouchi-Copelli neurons to analyze the information processing. The internal states of a Kinouchi-Copelli neuron is labeled by $s = 0,1,\cdots, n-1$ with $n=4$ here. The state $s=0$ represents the resting state and the state $s=1$ corresponds to the excited state. The other states $s=2, 3$ are refractory and are insensitive to stimulus. 

Time evolution of the Kinouchi-Copelli neuron is discrete. If the neuron is excited to the $s =1$ state at the current time step, it will evolve into the $s =2$ state in the next time step. And so on until the state $s = 3$ leads back to the $s=0$ resting state. During the numerical simulations, we update all neurons synchronously.

The simulation is based on the Monte-Carlo algorithm and the exposure time to collect the firing events is set to 1300 ms. During the first 200 ms, a constant stimulus stimulus is applied to the network, avoiding non-trivial trapping effects arisen from the initial conditions. Then, the following 800 ms is the pre-burn stage and the last 300 ms is the measurement stage. The firing events during the measurement stage are collected and analyzed. The numerical simulations are done on a PC cluster with 40 CPUs [Model: Intel(R) Xeon(R) CPU E5-2670 v2 at 2.50GHz]. For the largest size $L = 128$ of the Kinouchi-Copelli neuronal network, it takes about 8 hours to complete the 500 repeated simulations (using one core).

When analyzing the visual information contained in a natural image, the pixel value $S_i$ (between 0 and 255) is translated into equivalent stimulus by the activation probability
\begin{eqnarray}
P_i = 10 ^{-5(1-S_i/255)}.
\end{eqnarray}
Thus, the probability ranges between $10^{-5}$ (darkest $S_i=0$) and $1$ (brightest $S_i=255$) and the visual information of each pixels can be processed by the finite KCNN via numerical simulations.

\section{Statistical fluctuations}
\label{app2}
The critical point of a continuous phase transition is characterized by the onset of the maximum statistical fluctuations. Here, we introduce the fluctuations of the time-averaged activity,
\begin{equation}
  \Delta F =  (\langle F^{2} \rangle - \langle F \rangle ^{2})\cdot L^{2}
\end{equation}
where $\langle \cdot \rangle$ denotes the ensemble average. The ensemble size for the data points away from the critical point is set to 3000 while it is necessary to increase the ensemble size to 20000 in the vicinity of the critical point. The results are summarized in Figure S1.

\setcounter{figure}{0}
\makeatletter 
\renewcommand{\thefigure}{A\@arabic\c@figure}
\makeatother

\begin{figure}
\begin{center}
\includegraphics[width=\columnwidth]{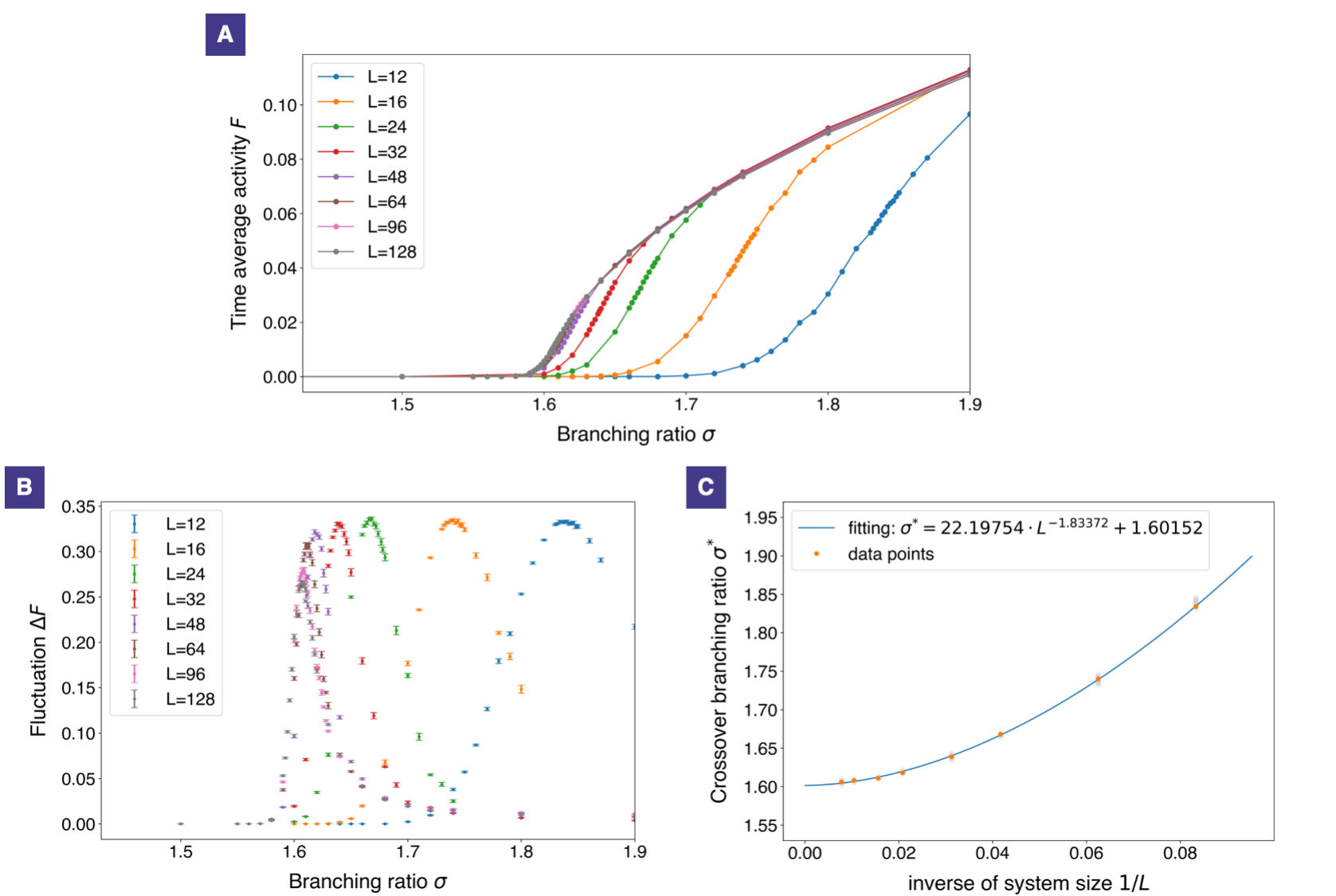}
\caption{\textbf{Time-averaged activity and its statistical fluctuations.} (A) The time-average activities for the KCNN of different sizes (from $L=12$ to $L=128$) are calculated versus the branching ratio. In the subcritical regime with smaller branching ratios, the activity reduces to zero in the absence of the external stimuli, corresponding to the quiescent phase. In the supercritical regime with larger branching ratios, spontaneous asynchronous firing occurs and the neuronal network is in the SAF phase. (B) To spot the critical regimes for the KCNN of different sizes, it is helpful to compute the statistical fluctuations of the time-averaged activity. The crossover branching ratio $\sigma^* = \sigma^*(L)$, defined as the onset where the strongest statistical fluctuations occur, moves to the smaller value as the system size grows. (C) Performing the finite-size scaling, the critical branching ratio $\sigma_c = 1.60152$ in the thermodynamic limit can be extracted with a non-trivial exponent $1.83372$.}
\end{center}
\end{figure}

\begin{figure}
\begin{center}
\includegraphics[width=\columnwidth]{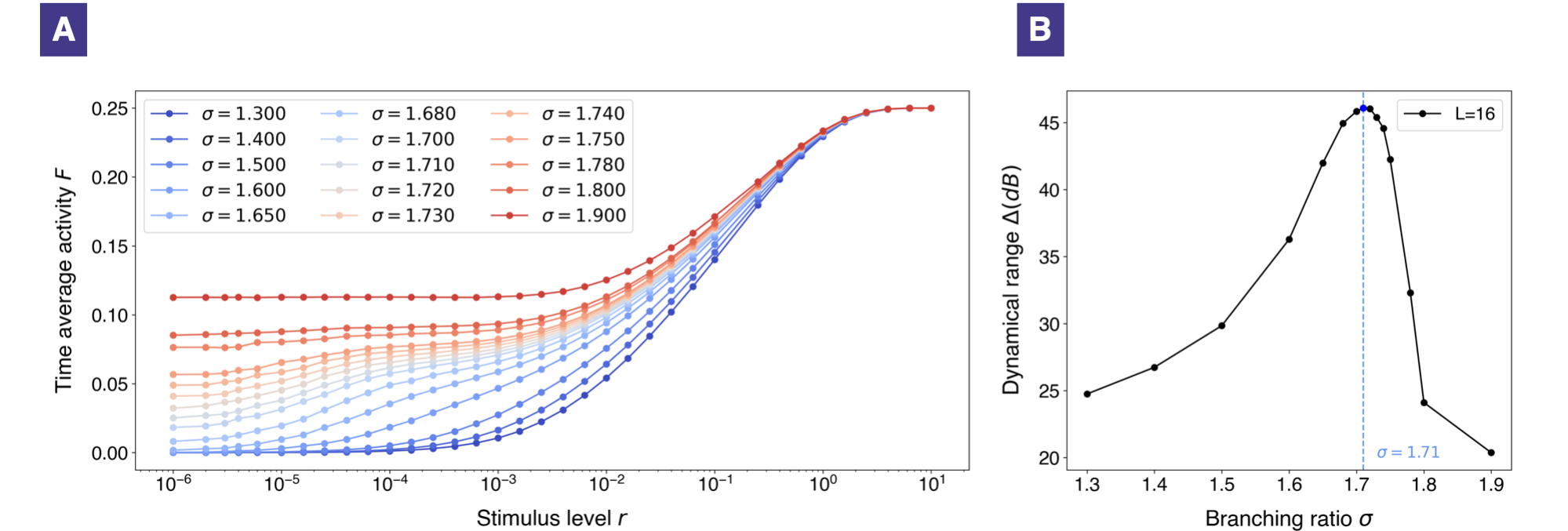}
\caption{\textbf{Response curve and its dynamical range.} (A) The response curve of the time-averaged activity $F=F(r)$ versus the external stimulus $r$ is computed at different branching ratios. The blue and red curves correspond to the quiescent (Q) and spontaneous asynchronous firing (SAF) phases respectively. (B) The dynamical range of the response curve can be calculated by the logarithmic ratio introduced in the Supplementary Information and reaches the peak around $\sigma_{\rm DR}=1.71$.}
\end{center}
\end{figure}

\section{Dynamical range}
\label{app3}

The activities of each neurons in the network can be described by the time-averaged activity $F$:
\begin{equation}
F = \frac{1}{T}\sum_{t}^{T}\rho_{t},
\end{equation}
where $\rho_{t}$ is the instantaneous neural activity of the network (the density of the active neuron at a given time $t$), and $T$ is a time window, which we take $T=300$ ms, for the measurement stage. The time-averaged activity $F$ in the subcritical regime always equals to zero, corresponding to the quiescent phase (Q phase). On the other hand, in the supercritical regime, a positive $F$ emerges and the neuronal network exhibits spontaneous asynchronous firing (SAF) behaviors, corresponding to the SAF phase.

In the presence of the external stimulus $r$, the time-averaged activity $F=F(r)$ can be computed in a similar fashion. As the strength of the stimulus grows, the activity goes from $F_0$ to the saturated value $F_{\rm max}$. The dynamical range, a measure of the sensitivity of the neuronal network, is defined by the logarithmic ratio,
\begin{equation}
\Delta = 10\log_{10}\frac{r_{0.9}}{r_{0.1}}
\end{equation}
where $r_{x}$ corresponds to the time-averaged activity $F_{x} = F_{0} + x(F_{\text{max}} - F_{0}), x \in [0,1]$. The results are summarized in Figure S2.

\section{Branching ratio for a finite KCNN}
\label{app4}

There are two ways for a neuron $i$ to get excited from the resting state. One is to be stimulated by a neighbor neuron $j$ which is in the excited state in the previous time step. We arrange synaptic weights within the network to be uniform so that each neuron can be equally stimulated by any of its neighbors. Therefore, the probability that an active adjacent neuron $j$ stimulates the neuron $i$ is given by a uniform synaptic weight $W_{ij}=W/4$. For a $L\times L$ square-lattice network, neurons inside the network have four neighbors, and neurons on the boundary have three (on the edge) or two (at the corner) neighbors. 

We define the local branching ratio $\sigma_{j}=\sum_{i}^{K_j}W_{ij}$ to be the average number of excitations created in the next time step by the $j$th neuron, where $K_j =4, 3, 2$ is the number of neighbors for the $j$th neuron. The branching ratio $\sigma$ is defined as the average over the local branching ratios $\sigma_j$ on the $L \times L$ neuronal network,
\begin{eqnarray}
\sigma=\frac{1}{L^2}\sum_{j} \sigma_{j} = \frac{1}{L^2} (4L^2-4L) \frac{W}{4} = \left(1-\frac{1}{L} \right) W
\end{eqnarray}
Thus, on a finite KCNN, the branching ratio $\sigma = (1-1/L) W$ is related to the synaptic weight $W$ with a finite-size correction.

\section{Branching ratio for a finite KCNN}
\label{app5}

There are two ways for a neuron $i$ to get excited from the resting state. One is to be stimulated by a neighbor neuron $j$ which is in the excited state in the previous time step. We arrange synaptic weights within the network to be uniform so that each neuron can be equally stimulated by any of its neighbors. Therefore, the probability that an active adjacent neuron $j$ stimulates the neuron $i$ is given by a uniform synaptic weight $W_{ij}=W/4$. For a $L\times L$ square-lattice network, neurons inside the network have four neighbors, and neurons on the boundary have three (on the edge) or two (at the corner) neighbors. 

We define the local branching ratio $\sigma_{j}=\sum_{i}^{K_j}W_{ij}$ to be the average number of excitations created in the next time step by the $j$th neuron, where $K_j =4, 3, 2$ is the number of neighbors for the $j$th neuron. The branching ratio $\sigma$ is defined as the average over the local branching ratios $\sigma_j$ on the $L \times L$ neuronal network,
\begin{eqnarray}
\sigma=\frac{1}{L^2}\sum_{j} \sigma_{j} = \frac{1}{L^2} (4L^2-4L) \frac{W}{4} = \left(1-\frac{1}{L} \right) W
\end{eqnarray}
Thus, on a finite KCNN, the branching ratio $\sigma = (1-1/L) W$ is related to the synaptic weight $W$ with a finite-size correction.

\section{Computation of mutual information}
\label{app6}
In this paper, we use a $512\times512$ pixel image `butterfly' as our source of nature image and cut it randomly into an ensemble of 5000 pieces of $16\times16$ images. We continuously stimulate the neuronal network with the same image within a trial of 1000 time steps. After doing the same experiment for all images in the ensemble, we will get a joint probability distribution $P(X_i,Y_i)$ of the input stimulus $X_i$ and the time average activity $Y_i$ for each neuron $i$. We map the input stimulus $X_i \rightarrow X_i^{'}=X_i-\frac{1}{256}\sum_j^{256} X_j$ to consider the spatial correlation of image inputs and get another joint probability distribution $P(X_i^{'},Y_i)$. From this new probability distribution, we can calculate the mutual information between the zero-sum filtered input $X_i^{'}$ and the time average activity $Y_i$ for each neuron $i$. Last, we take the average of mutual information over all neurons in the network and use the average branching ratio as our order parameter to study the information processing in this network. We also demonstrate how mutual information will be if we change the size of the network to be $12\times12$ or change the source image to be `Lenna' and `birds' as shown in Figure S3.

\begin{figure}
\begin{center}
\includegraphics[width=\columnwidth]{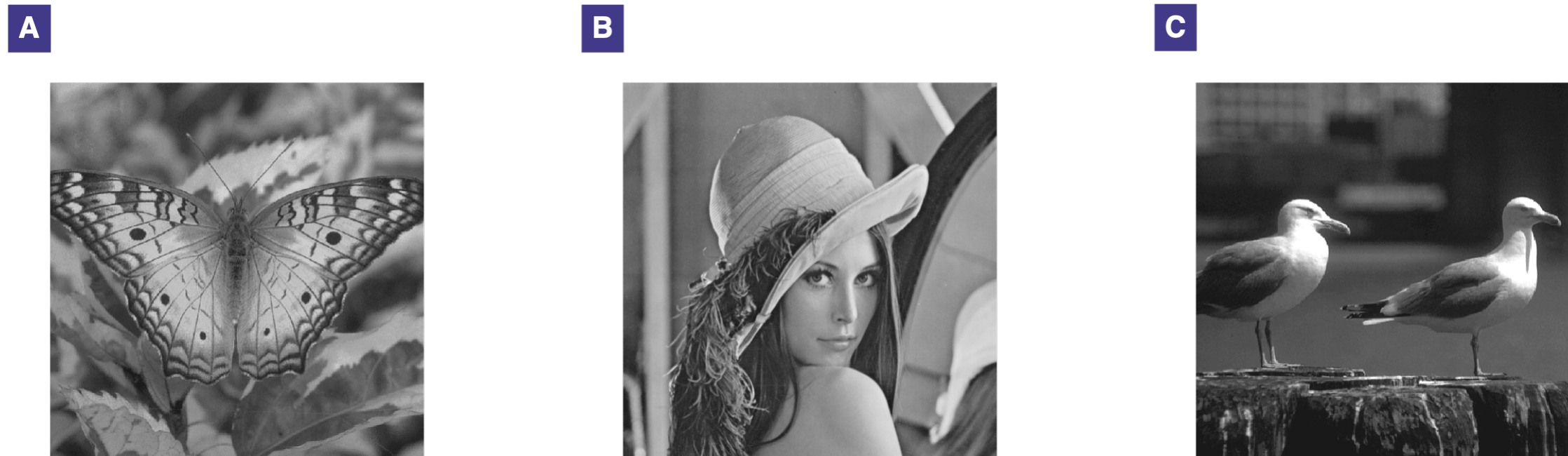}
\caption{\textbf{Different natural images.} (A) butterfly. (B) Lena. (C) birds. }
\end{center}
\end{figure}


\end{document}